\documentclass[useAMS,usenatbib]{mn2e}

\usepackage{graphicx}

\title[New Pulsating White Dwarfs in Cataclysmic Variables]{New Pulsating White Dwarfs in Cataclysmic Variables\thanks{Based on observations made with the Nordic Optical Telescope, operated on the island of La Palma jointly by Denmark, Finland, Iceland, Norway, and Sweden, in the Spanish Observatorio del Roque de los Muchachos of the Instituto de Astrofisica de Canarias.}}
\author[R. Nilsson et al.]
{R. Nilsson$^{1}$\thanks{E-mail: ricky@astro.lu.se}, H. Uthas$^{1}$, M. Ytre-Eide$^{2}$, J-E. Solheim$^{2}$, B. Warner$^{3}$ \\
$^{1}$Lund Observatory, Box 43, SE-22100 Lund, Sweden\\
$^{2}$Institute of Theoretical Astrophysics, Box 1029 Blindern, N-0315 Oslo, Norway\\
$^{3}$Department of Astronomy, University of Cape Town, Rondebosch 7700, S.\ Africa}

\begin{document}

\date{Accepted 2006 May 10. Received 2006 May 1; in original form 2006 March 22}

\pagerange{\pageref{firstpage}--\pageref{lastpage}} \pubyear{2006}

\maketitle

\label{firstpage}

\begin{abstract}
The number of discovered non-radially pulsating white dwarfs (WDs) in cataclysmic variables (CVs) is increasing rapidly by the aid of the Sloan Digital Sky Survey (SDSS). We performed photometric observations of two additional objects, SDSS$\:$J133941.11+484727.5 (SDSS$\:$1339), independently discovered as a pulsator by G{\"a}nsicke et al., and SDSS$\:$J151413.72+454911.9, which we identified as a CV/ZZ Ceti hybrid. In this Letter we present the results of the remote observations of these targets performed with the Nordic Optical Telescope (NOT) during the Nordic-Baltic Research School at Mol\.etai Observatory, and follow-up observations executed by NOT in service mode. We also present 3 candidates we found to be non-pulsating.

The results of our observations show that the main pulsation frequencies agree with those found in previous CV/ZZ Ceti hybrids, but specifically for SDSS$\:$1339 the principal period differs slightly between individual observations and also from the recent independent observation by G{\"a}nsicke et al. Analysis of SDSS colour data for the small sample of pulsating and non-pulsating CV/ZZ Ceti hybrids found so far, seems to indicate that the $r-i$ colour could be a good marker for the instability strip of this class of pulsating WDs. \end{abstract}


\section{Introduction}
Non-radially pulsating white dwarfs (WDs) of DA type (DAV), known as ZZ Ceti stars, have up to recently almost exclusively been found as single isolated objects. However, the start of the Sloan Digital Sky Survey \citep[SDSS:][]{b8} in 2000 has enabled the discovery of several cataclysmic variables (CVs) harbouring pulsating primaries. Spectra of faint CVs obtained from SDSS have revealed a number of low-mass transfer rate dwarf nova systems with faint accretion discs, where light from the WD dominates the optical flux, hence allowing us to study low-amplitude modulations in the light-curve, induced by pulsations of the WD. Thus far, 10 such CV/ZZ Ceti hybrids have been found; the most recent one being SDSS$\:$J133941.11+484727.5 (henceforth SDSS$\:$1339) as described by \citet{b5}.

The pulsation frequencies observed in ZZ Ceti stars are often linear combinations of the main pulsation frequency together with eigenfrequencies of other principal driving modes, possibly described by a general numerical formula  \citep[see e.g.][]{b18}. For the limited number of CV/ZZ hybrid systems identified so far, the resonance condition appears to be slightly different from single ZZ Ceti stars. 

An accreting WD can be quite unlike an isolated WD, after having undergone about $10^9$ yr of accretion and several nova eruptions. Since the interior structure might be different we also expect the fingerprint frequencies of eigenmode pulsations to be different. Asteroseismological analysis of non-radially pulsating WDs as primaries in CVs can give us important information about structure, composition and evolution of the WD as well as the accretion process, e.g.\ help us determine the mass of the primary and the accreted hydrogen layer, and improve our models of classical novae. The pulsation eigenfunctions could be affected by e.g.\ accretion-induced spherical asymmetries (due to equatorial band accretion in the low magnetic field WD primary), rapid rotation (due to angular momentum transfer from accreted material), and temperature fluctuations (due to outburst on long time-scales and accretion variations on shorter time-scales). Furthermore, clumpy and non-smooth accretion flow, indicated by flickering, may induce random phase changes as it continuously excites oscillations. As the WD evolves toward the cool limit of the instability strip for DAVs the pulsation spectra becomes increasingly complex and unstable, with amplitudes changing considerably on time-scales of only months \citep[][and references therein]{b14}.
\begin{table*}
 \centering
 \begin{minipage}{120mm}
  \caption{Observing log for the remote NOT observations from Mol\.etai of all five candidates and the fast-track service observations of SDSS$\:$1339 and SDSS$\:$1514.}
  \begin{tabular}{@{}llllllll@{}}
  \hline
  Date  &   Object & V & Start & Length &Cycle &  Points & Observers\\    
        &          &               &[\textsc{ut}]&  [s]    &   [s] &        &          \\
  \hline
  2005-08-14 &    SDSS$\:$1339 & 18   & 22:17     &2350 & 40 &  61 & HU\footnotemark{}
 \& BIV\footnotemark{}  
  \\
  2005-08-15 &    SDSS$\:$1501 & 19.5 & 21:58  &2800     & 40 &  70 & OS\footnotemark{}
 \& AVBH\footnotemark{}
 \\
  2005-08-15 &    SDSS$\:$1610 & 19.5 & 23:28      &2040& 40 &  51 & SAGS\footnotemark{}
 \& SM\footnotemark{} \\
  2005-08-16 &    SDSS$\:$1507 & 18   & 20:57      &4400& 30 & 147 & MYE\footnotemark{}
 \& ES\footnotemark{}
  \\
  2005-08-16 &    SDSS$\:$1514 & 19.5 & 22:37       &4080& 40 & 102 & RN\footnotemark{}
 \& EP\footnotemark{} 
   \\
  2005-08-28 &    SDSS$\:$1339 & 18   & 21:04      & 3263&32.6 & 100 & GM\footnotemark{}
 \& RK\footnotemark{}
  \\
  2005-08-28 &    SDSS$\:$1514 & 19.5 &  22:01       &6650& 33.2 & 200 & GM \& RK
  \\
 \hline
 \end{tabular}
 
Observer key: $^{1}$H.~Uthas, $^{2}$B.~I.~Vik, $^{3}$O.~Smirnova, $^{4}$A.~V.~B.~Hansen, $^{5}$S.~A.~G.~de Sousa, $^{6}$S.~Mikolaitis, $^{7}$M.~Ytre-Eide, $^{8}$E.~Stasiukaitis, $^{9}$R.~Nilsson, $^{10}$E.~Pukartas, $^{11}$G.~Micheva (NOT), $^{12}$R.~Karjalainen (NOT). 
 \end{minipage}
 \end{table*}
In this Letter we present the independent discovery of pulsations in SDSS$\:$1339 and one additional CV/ZZ Ceti hybrid candidate, SDSS$\:$J151413.72+454911.9 (henceforth SDSS$\:$1514), found during remote observations with the Nordic Optical Telescope (NOT) in 2005 August, while attending the Nordic-Baltic research school at Mol\.etai Observatory (Lithuania), and also examine the results of follow-up observations with NOT later that month. We also briefly present the results of observations of 3 other candidates: SDSS $\:$J150137.22+550123.4 (called SDSS$\:$1501), SDSS$\:$J150722.33+523039.8 (SDSS$\:$1507), and SDSS$\:$J161030.35+445901.7 (SDSS$\:$1610) for which we did not detect pulsations.
\vspace{-3mm}
\section{Observations}
Remote NOT observations of five CVs found in the SDSS were performed from Mol\.etai Observatory in Lithuania in 2005 August. For two targets, SDSS$\:$1339 and SDSS$\:$1514 additional observations were performed by NOT in service mode (Fast-Track Service Program) on August 28.
\vspace{-3mm}
\subsection{Target selection}
The targets were selected from a list of possible Northern Hemisphere candidates for pulsating CVs by Brian Warner and Patrick A.~Woudt as an extension of their work in the Southern Hemisphere on detection of pulsations in CVs \citep{b18,b19}.
The selection of targets was based on their spectra from SDSS showing clear signs of absorption lines from the WD primaries, thus indicating a low relative flux contribution from the accretion disc and secondary star.
\vspace{-3mm}
\subsection{Instrumentation}
The observations were conducted using the Andalucia Faint Object Spectrograph and Camera (ALFOSC). Applying the filter W92, which has a full width at half-maximum (FWHM) of 275 nm centred at 550 nm, we were able to gather a fair amount of flux from the relatively faint targets and at the same time minimize the contribution from the infrared sky background.  

The NOT data acquisition was controlled remotely by using the software interface \textsc{tcpcom} in a mode called `windowed fast photometry mode'. The light curves, including both the raw data and the sky-subtracted data, were displayed in real time using the program \textsc{rtp} (real time processing).
\vspace{-3mm}
\subsection{Observations of SDSS$\:$1339 and SDSS$\:$1514}
The remote observations of SDSS$\:$1339 were performed using seven readout windows (one target, four comparison stars and two sky windows). Exposure time for each frame was approximately 33~s and readout time approximately 7~s, adding up to a total cycle time of 40~s. It was observed under good conditions, but the target was somewhat close to the horizon and the moon was 71 per cent illuminated. SDSS$\:$1514 was observed under similar condition and a moon of 89 per cent. See Table 1 for additional information regarding the observations. 

\subsection{Fast-Track Service Program}
The Fast-Track Service Program gives the possibility to propose a short observing program of 4 hrs which can be conducted on short notice by NOT. %The instrument choice is limited to ALFOSC, NOTCam or StanCam and target acquisition and calibrations such as observations of standard stars must be conducted within the 4 hours of observation time. 

Observations of SDSS$\:$1339 and SDSS$\:$1514 were conducted on 2005 August 28, using ALFOSC. An exposure time of 25~s was set for both targets and the number of frames to 100 for SDSS$\:$1339 and 200 for SDSS$\:$1514.  An observing window of 100 ${\times}$ 100 arcsec (524 ${\times}$ 524 pixels) and a 2 ${\times}$ 2 pixel binning readout was used, and the total cycle time became approximately 33~s.
\vspace{-3mm}
\section{Data reduction and analysis}
Reduction of the initial raw data proceeded along the usual basic steps, viz. flat-fielding, background subtraction, aperture photometry, air-mass compensation and Fourier transform (FT) analysis to search for possible periodic modulations caused by pulsations. As these objects are mass-transferring, we also expect periodic variations on longer time-scales related to their orbital periods. Quasi-periodic variations on shorter time-scales related to flickering may also be observed. 

%The data files obtained at Mol\.etai were processed with the \emph{rtp} program and \emph{xqed}, giving us the flat-field corrected and background subtracted data.

%Images from the service night were somewhat different and were also treated differently in the reduction process. \emph{Imcat} software was used for flat-fielding, and \emph{SExtractor} for source extraction and background subtraction. The output data-files were read into \emph{R} where we performed source/comparison division.  The program \emph{qfitsview} developed by R. Janulis was also used alternatively for the data reduction.

For detection of a pulsation we require a peak on the order of 3$\sigma$ above the nearby noise in the FT. Because our observations were quite short and the objects may show quasi-periodic variations, a second run was necessary for a safe detection. Images from the service night observations were processed with \textsc{SExtractor} for optimal source extraction and background subtraction, giving lower overall noise.
\vspace{-3mm}
\section{Results}
\subsection{SDSS$\:$1339}

\begin{figure}
 \vspace{0pt}
 \includegraphics[scale=1.0]{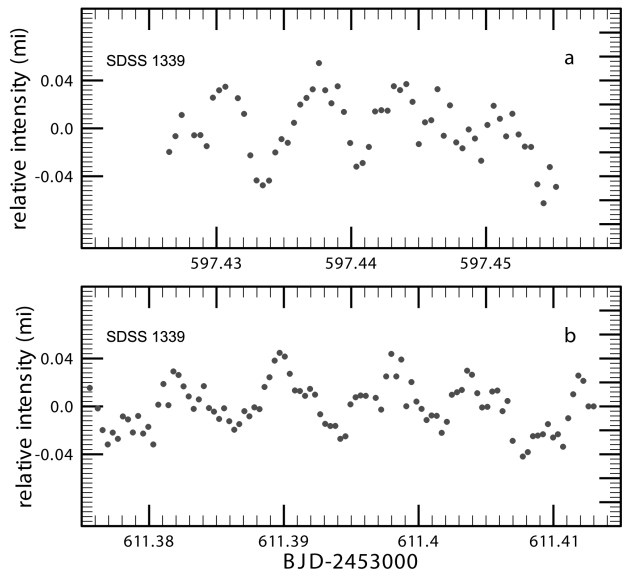}
 \caption{Light-curves of SDSS$\:$1339 obtained on (a) August 14 and (b) August 28.}
\end{figure}

In the light curves of SDSS$\:$1339 displayed in Fig.~1, we clearly observe four pulses the first run and five pulses the second run. The pulses in the second run are more triangular than in the first. In the FT (Fig.~2) we find a significant peak at 1.68 mHz (or 598 s) with an amplitude of 25 mma in the first run, and at 1.52 mHz (659 s) with an amplitude of 20 mma in the second run. There are also other peaks below the significant detection limit. None of these repeat in both runs, and may be due to quasi-periodic flickering. 

\begin{figure}
 \vspace{0pt}
 \includegraphics[scale=1.0]{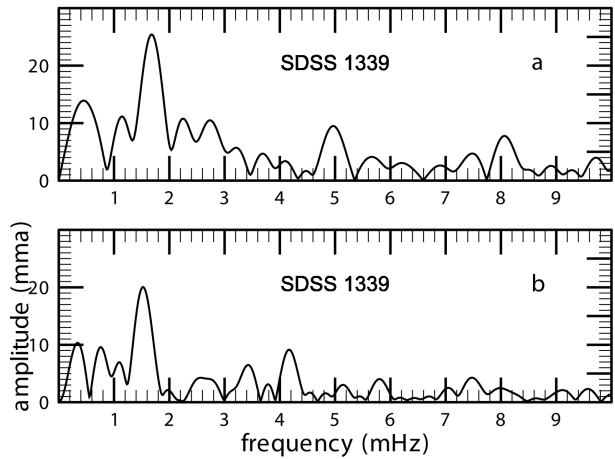}
 \caption{Fourier transforms of the SDSS$\:$1339 light-curves from the two runs (a) August 14 and (b) August 28.}
\end{figure}

\vspace{-3mm}
\subsection {SDSS$\:$1514}
This object is fainter, and was observed the first time only 3~d from full moon, and the
second time with variable sky background. In Fig.~3 we show the FTs for the two runs, individually and combined. In the first run, on 2005 August 16, we find a significant peak, with an amplitude above the False Alarm probability (FALSE = 1/100) \citep{b22}, at 1.79 mHz (557 s). This repeats in the second run, but is not significantly higher than the noise, with an amplitude of only 7.1 mma. This may be interpreted as if the possible pulse has disappeared. However, we have performed the data reduction with different methods of subtracting the background sky, and also with division/no division of the light curve of a comparison star. The noise pattern changes between the different reductions, but one peak at about 1.8 mHz is always present. We interpret this as if the pulsation is real and present with low amplitude. The combined FT gives a peak at 1.79 mHz or a period of 559.3 s. The final result is given in Table 2.

In Fig.~4 we show the average pulse shape of the period of 559.3 s.  Each phase point is
an average of 18 periods and smoothed over approximately 100 s. The pulse shape is nearly
sinusoidal with a slower rise and faster decline. 

\begin{figure}
 \vspace{0pt}
 \includegraphics[scale=1.0]{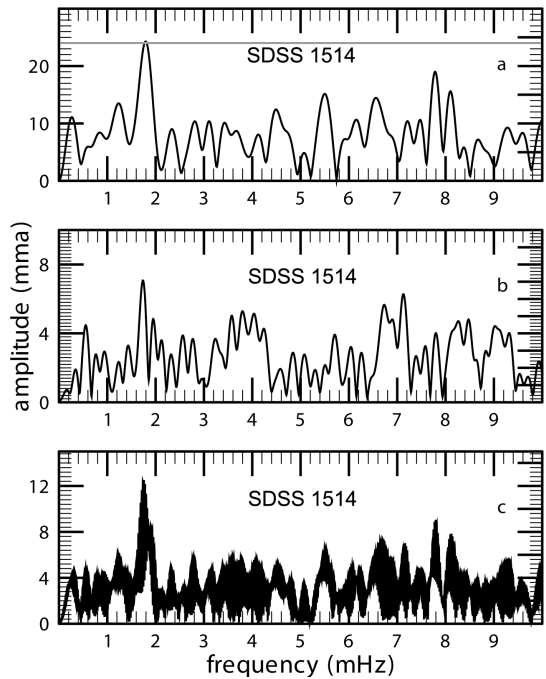}
 \caption{Fourier transforms of the SDSS$\:$1514 light curves from the two runs separately (a) August 16 with False Alarm (FALSE = 1/100) level indicated;  (b) August 28 and (c) combined. Only the modulation near 1.8 mHz is present in both runs. In the second run we find that the noise pattern changes when we use different ways of sky subtraction, while the 1.8-mHz peak is always present.}
\end{figure}

\begin{figure}
 \vspace{0pt}
 \includegraphics[scale=1.0]{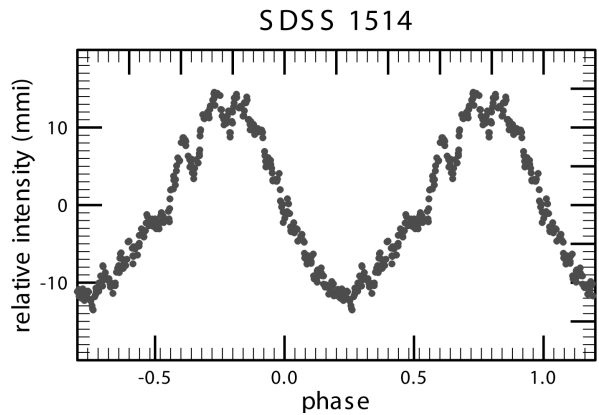}
 \caption{The pulse shape of the strongest modulation in SDSS$\:$1514.}
\end{figure}

\vspace{-3mm}
\subsection {The non-pulsators}

%In figure 5 we show the FT of the non-pulsators. 

SDSS$\:$1501 is an eclipsing binary, having a deep minimum with no significant brightness variations outside the eclipse, and a light curve similar to that of UX~UMa. The FT of the light curve outside the eclipse showed no significant peaks.

SDSS$\:$1507 also showed a deep eclipse, and considerable brightness variations outside the eclipse. Its light-curve is quite similar to Z~Cha, as both eclipses of the WD and the bright spot are visible, in addition to a strong reflection effect. The FT in this case shows many peaks below 3 mHz, but they are most likely due to quasi-periodic modulations, which are often observed in such CVs.

Finally, SDSS$\:$1610 showed a light-curve without any strong modulations.

%\begin{figure}
% \vspace{0pt}
% \includegraphics[scale=0.67]{Fig5A.eps}
 %\caption{Fourier transforms of the light-curves from the 3 candidates which show no clear
%sign of pulsations near 2 mHz.}
%\end{figure}

\begin{figure}
 \vspace{0pt}
 \includegraphics[scale=1.0]{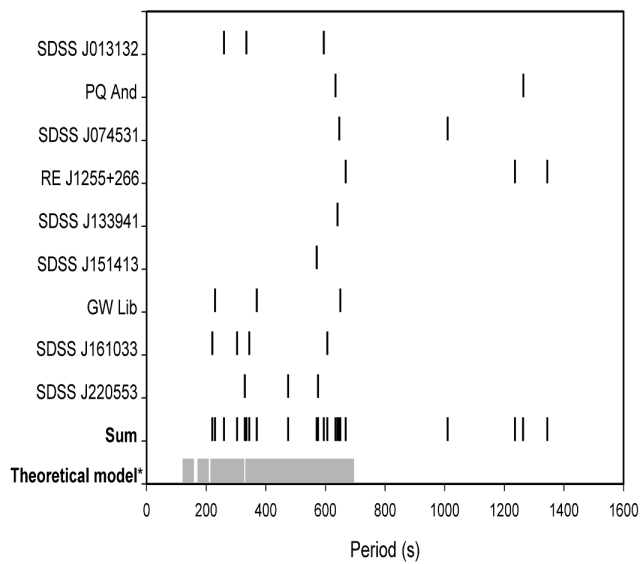}
 \caption{Overview of observed pulsation spectra for all known CV/ZZ Ceti hybrids compared with a theoretical period range ($^{*}$adapted from model calculations of GW Lib by Towsley, Arras \& Bildsten 2004).}
\end{figure}
\vspace{-4mm}
\section {Discussion}
Fig.~5 portrays the pulsation spectra of all known CV/ZZ Ceti hybrids and compares the total of all observed periods with a theoretical period range for accreting ZZ Ceti stars \citep[based on results from a model calculation of GW Lib by][]{b12}. The two significant peaks observed in SDSS$\:$1339 and SDSS$\:$1514 (Table 2) match some of the main periods found in previous objects identified as CV/ZZ Ceti hybrids, clustering around 600~s.

\begin{table}
  \caption{Identified frequencies in the Fourier data of SDSS$\:$1339 
  and SDSS$\:$1514.}
  \begin{tabular}[h!]{@{}llll@{}}
  \hline
Object            & Date       &  Frequency [${\mu}$Hz] & Amplitude [mma]\\    
  \hline
SDSS$\:$1339      & August 14  & 1678      & 25  \\
SDSS$\:$1339      & August 28  & 1517      & 20  \\
SDSS$\:$1339      & Combined  & 1587     & 20     \\
SDSS$\:$1514      & August 16  & 1794      & 25 \\
SDSS$\:$1514     & August 28  & 1735  & 7  \\
SDSS$\:$1514      & Combined  & 1788   & 13 \\
\hline
\end{tabular}
\end{table}

If we combine the two runs on SDSS$\:$1339 we get a peak at 1.517~mHz or 630~s.  However, the accuracy in the frequency determination of each individual observation suggests that the frequency actually is changing. The period obtained in the first run is noticeably shorter than the period of 642 seconds observed by \citet{b5} during measurements in 2005 April (even taking into account our much shorter run time), while our second observation gave a distinctly longer period. Although one could suspect such a difference in the measured main pulsation period to be caused by effects of flickering or perhaps excitation of a nearby mode, one might also argue that this is a real drift of the main pulsation mode. Intermittent onset of mass transfer caused by thermal instabilities in the accretion disc can occur on a regular basis once every few months and last for about a week, eventually leading to a build-up and sedimentation of matter, which will cause compressional heating of the WD and consequently change the eigenmode frequencies. A decrease in the observed pulsation period might well be a sign of an approaching dwarf novae (DN) outburst, as heating of the WD core through material compression usually takes place just prior to unstable ignition \citep{b13}. Another contributing factor could be influence from simmering nuclear burning affecting mode periods and period spacings \citep{b2}. GW Lib can serve as an illustration of the opposite effect where the mode frequencies drift because of cooling of freshly accreted material after a DN outburst. Calculations by \citet{b12} match the frequency decrease of the periodicity near 646 s observed for GW Lib by \citet{b14} quite well. The observed variation, $\dot{\nu}=\dot{\omega}/2\pi=-10^{-11}\: \textrm{Hz s}^{-1}$, at the specific frequency 1/646 Hz, gives an increase in the mode period by roughly 10 s per month \citep{b12,b13}. For an isolated ZZ Ceti the period change is directly related to the evolutionary time-scale, thus the periods are very stable. Another example of observed rapid period variability is the pulsating central WD of the planetary nebula NGC$\:$246 which has shown a change of 130 $\mathrm{\mu}$Hz in just 3~d, perhaps implying that it is a member of a binary system \citep{b4}.

\begin{figure}
 \vspace{0pt}
 \includegraphics[scale=0.6]{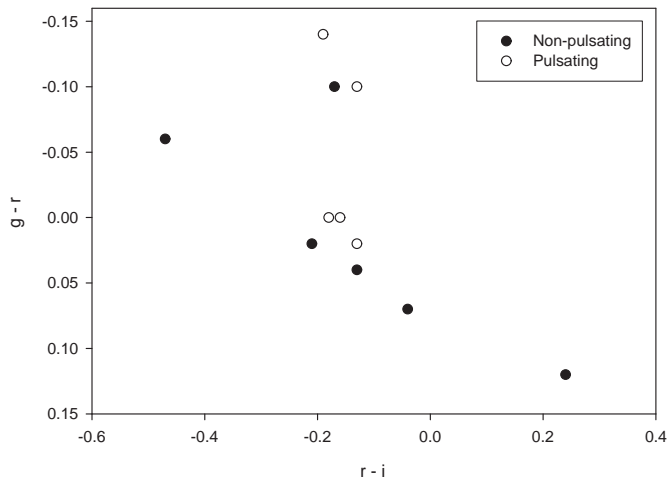}
 \caption{Colour-colour diagram displaying the positions of pulsating and non-pulsating WDs in CV systems, from magnitude data presented in the SDSS data releases \citep{b8,b9,b10,b11}. The $g$, $r$ and $i$ filters are given by the Gunn pass-bands centred around 520, 670 and 790 nm, respectively.}
\end{figure}

The pulse shape of the main pulsation in SDSS$\:$1514 is nearly sinusoidal. Deviations from the sinusoidal shape and linear combinations of frequencies in the Fourier spectrum may be a result of non-linear effects, perhaps due to perturbations in the convection zone \citep[see][for review on non-linear processes]{b21}. Due to the non-pulsating contribution by the accretion disc, it is difficult to determine the actual pulsation amplitude in hybrid systems. Amplitude variations on time-scales of just days is a frequent feature of WD pulsators near the red edge of the instability strip, and should be common also among CV/ZZ Ceti hybrids, so the observed change in pulsation amplitude of SDSS$\:$1514 is not surprising. Had this object been observed solely on the 28th it would probably have been declared a non-pulsator and dropped for further investigation\footnote[1]{Observations of SDSS$\:$1514 on May 5, 2006 did not show any significant pulsations, further indicating that the object is an unstable pulsator. (Added in proof.)}. Other seemingly non-pulsators may only temporarily be in a low amplitude state due to variation in accretion or a position near either edge of the instability strip, thus it is important to get good signal-to-noise ratios in the observations.

From the performed observations and the analysis that followed we conclude that SDSS 1339 show clear signs of harbouring a non-radially pulsating WD, consistent with observations by \citet{b5}, and that SDSS 1514 shows strong indications of being a pulsator. Many more CV/ZZ Ceti hybrids are undoubtedly out there. In fact, one might speculate that the mass-transfer on to the WD primary can be an excitation mechanism which channels energy into pulsations, thus there could be an increased probability of non-radial pulsations in accreting WDs compared to isolated WDs. Possibly the $r-i$ colour interval in Fig.~6, where several of the pulsators reside, may be an indication of the pulsation strip for this type of accreting WD. Future observations will add to this currently small sample and hopefully reveal more explicitly if this colour interval is a good marker.
\vspace{-4mm}
\section*{Acknowledgments}
We would like to thank Hans Kjeldsen for all the help in preparing both the remote observations from Mol\.etai and the observing proposal for the follow-up observations, and Patrick A.\ Woudt for clever selection of new CV/ZZ Ceti candidates that grew from aspirants into (more or less) confirmed hybrids during the research course at Mol\.etai. In addition, credit goes to all students and assistants who helped out during the observations, and to Paula Szkody who deserves recognition for providing support and valuable information subsequent to the observations.

The data presented here have been taken using ALFOSC, which is owned by the Instituto de Astrofisica de Andalucia (IAA) and operated at the Nordic Optical Telescope under agreement between IAA and the NBIfAFG of the Astronomical Observatory of Copenhagen.

\appendix

\bsp

\label{lastpage}

\end{document}